\def\real{I \!\! R}
\def\vect{\underline}
\def\sgn{\hbox{sgn}}
\def\card{\hbox{card}}
\def\definition{\> {\buildrel \hbox{\tiny def} \over = } \>}
\def\eps{\varepsilon}
\def\mypicture#1{
      \centerline{\parbox{159mm}{\epsfxsize=159mm\epsffile{#1}}}
      \vskip0mm}
\def\reference#1#2#3#4{\vskip1ex\noindent\parbox{5mm}{\rightline{[#1]\ \ }}
\vtop{\hsize=0.95\hsize\noindent #2, {\it ``#3''}, #4}}

\documentstyle[12pt]{article}

\input{epsf}

\begin{document}

\baselineskip=24pt

\ \bigskip\bigskip\bigskip\bigskip

\centerline{\LARGE A polynomial training algorithm for}
\centerline{\LARGE calculating perceptrons of optimal stability}

\centerline{\large (published in J.\ Phys.\ A {\bf 28}(8), p.\ 2173--2181,
1995)}

\bigskip\bigskip\bigskip\bigskip

\baselineskip=18pt

\centerline{\large Jorg Imhoff}
\centerline{\large Universit\"at Heidelberg, Institut f\"ur
Theoretische Physik,}
\centerline{\large Philosophenweg 19, D-69120 Heidelberg}
\centerline{(e-mail: imhoff@hybrid.tphys.uni-heidelberg.de)}

\bigskip\bigskip\bigskip\bigskip\bigskip\bigskip\bigskip\bigskip

\parskip=3pt plus 1pt
\baselineskip=18pt

\centerline{\underbar{\bf Abstract}}

\bigskip

\centerline{\hbox{\vbox{\hsize=0.87\hsize
Recomi (\underbar{Re}peated \underbar{co}rrelation \underbar{m}atrix
\underbar{i}nversion) is a polynomially fast algorithm for searching optimally
stable solutions of the perceptron learning problem. For random unbiased and
biased patterns it is shown that the algorithm is able to find optimal
solutions, if any exist, in at worst ${\cal O}(N^4)$
floating point operations. Even beyond the critical storage capacity $\alpha_c$
the algorithm is able to find locally stable solutions (with negative
stability) at the same speed. There are no divergent time scales in the
learning process. A full proof of convergence cannot yet be given, only major
constituents of a proof are shown.
}}}

\newpage

Spin glass models of neural networks and their application as an associative
memory have been of great interest in the last years [1-10]. One major issue of
the field is the question of training networks, that is the construction of a
synaptic matrix in order to store given information. In this paper I am going
to present a training algorithm that is able to find solutions of the
perceptron problem of optimal stability in finite time. Unlike other
algorithms, as Minover presented by Krauth and M\'ezard [5] or AdaTron by
Anlauf and Biehl [6], this algorithm does not only approximate optimal
solutions but actually finds them. Furthermore, there are no divergent
timescales in the solution of the problem. Minover and AdaTron both have
diverging training times as the critical storage capacity $\alpha_c$ is
approached [6,7], whereas this algorithm does not. Therefore it can also be
used beyond $\alpha_c$ in the region of broken replica symmetry, where it finds
local optima of negative stability. A similar algorithm was proposed by Ruj\'an
[8], which also finds optimal perceptrons in finite time, but cannot advance
beyond $\alpha_c$.

Like the pseudo-inverse solution of the perceptron problem [9,10] this
algorithm uses inversion of pattern correlation matrices for searching
(optimal) perceptron couplings. As matrix inversion has to be done repeatedly,
the algorithm was called Recomi --- \underbar{Re}peated \underbar{co}rrelation
\underbar{m}atrix \underbar{i}nversion. As was shown by Opper [7] the problem
of finding an optimal perceptron is the problem of finding the subset of
embedded training patterns with minimal local fields. Recomi is able to find
this subset of patterns iteratively in finite time. The coupling vector is then
just the pseudo-inverse of the respective pattern correlation matrix.

I consider a network of $N+1$ neurons $S_i=\pm 1$, $i=1,\ldots,N+1$, coupled
through synaptic efficacies $J_{ij}$ (without taking self couplings into
account, i.e. $J_{ii}=0 \quad\forall i$). The dynamics of the system is taken
to be a simple zero-temperature Monte Carlo process:

\begin{equation}
S_i(t+1) = \sgn \left( \sum_{j(\neq i)} J_{ij} S_j(t) \right)
\end{equation}

The purpose of perceptron training algorithms is to find couplings $J_{ij}$
such that $p$ patterns $\vect\eta^\mu=(\eta^\mu_1, \ldots,
\eta^\mu_{N+1})^T$, $\eta^\mu_i=\pm 1$, $\mu=1,\ldots,p$, become fixed points
of the dynamics. That is

\begin{equation}
\eta^\mu_i \sum_{j(\neq i)} J_{ij} \eta^\mu_j \geq \kappa > 0,
\qquad i=1,\ldots,N+1; \qquad \mu=1,\ldots,p.
\end{equation}

The problem can be reformulated by looking at the single neurons (or simple
perceptrons) of the network, e.g.\ neuron $N+1$. With

\begin{equation}
\xi^\mu_i \definition \eta^\mu_{N+1} \eta^\mu_i,
\qquad i=1,\ldots,N; \qquad \mu=1,\ldots,p
\label{xi}
\end{equation}

\noindent one now has to find couplings $J_i$, $i=1,\ldots,N$, such that

\begin{equation}
h_\mu = \sum_i J_i \xi^\mu_i \geq \kappa >0, \qquad \mu=1,\ldots,p.
\end{equation}

\noindent If the norm of $\vect J$ is fixed, e.g. $|\vect J|=1$, it is
possible to define what is meant by ``optimal solutions'' of the given
problem:

\begin{equation}
\hbox{maximize\ \ } \kappa = \min_\mu \{ h_\mu \}
\hbox{\ \ under the constraint\ \ } \left| \vect J \right| = 1
\label{optimization-problem}
\end{equation}

With maximal $\kappa$ one expects to have maximum stability against input
noise, i.e.\ maximal basins of attraction in a network of neurons.

{}From the point of view of mathematical optimization it suitable to
reformulate the problem. With $ \vect J \longrightarrow \vect J / |\kappa| $
one gets an equivalent formulation of problem (\ref{optimization-problem}):
\begin{eqnarray}
\hbox{minimize\ \ } \left| \vect J \right| &
\hbox{under the constraints} \quad h_\mu = \vect J^T \vect\xi^\mu
\geq +1 \quad \forall\mu & \hbox{(for $\kappa>0$)}
\label{optim-prob-pos-kappa} \\
\hbox{maximize\ \ } \left| \vect J
\right| & \hbox{under the constraints} \quad h_\mu = \vect J^T
\vect\xi^\mu \geq -1 \quad \forall\mu & \hbox{(for
$\kappa<0$)} \label{optim-prob-neg-kappa}
\end{eqnarray}

I will use this formulation of the problem later in this article. Applying the
Kuhn-Tucker theorem of optimization theory [11] it can be shown [7] (see also
[6]) that an optimal solution, for $\kappa>0$, can always be written in the
form

\begin{equation}
\vect J = \sum_{\mu\in\Gamma} x_\mu \vect\xi^\mu,
\quad \hbox{where\ \ } x_\mu \geq 0 \quad \forall \mu\in\Gamma
\label{positive-embeddings}
\end{equation}
\noindent with
\begin{equation}
h_\mu = \vect J^T \vect\xi^\mu
   \cases{ = \kappa & $\mu\in\Gamma$ \cr
           > \kappa & else \cr
         }
\label{and-the-fields}
\end{equation}

For $\kappa<0$ the same argument holds for all local optima, but with $x_\mu
\leq 0$ $\forall \mu\in\Gamma$. $\Gamma$ is the set of ``embedded'' patterns,
$\Gamma \subseteq \{1,\ldots,p\}$. The $x_\mu$ are called the embedding
strengths of solution $\vect J$. Anlauf and
Biehl have also shown [6] that for $\kappa>0$
this solution is unique (which is in general not the case for $\kappa<0$).
I.e.\ two solutions $\vect J$ and $\vect J^\ast$ of the form
(\ref{positive-embeddings})(\ref{and-the-fields}) are always identical $\vect J
\equiv \vect J^\ast$. Note that if $\{ \xi_\mu | \: \mu\in\Gamma \}$ is a set
of linearly independent vectors --- e.g.\ if the patterns are in general
position and $\card(\Gamma) \leq N$ --- the choice of the $x_\mu$ is
unambiguous. On the other hand, if one has a solution of the form
(\ref{positive-embeddings})(\ref{and-the-fields}) it must be the global optimum
of the problem.

In the following sections I am going to describe the Recomi algorithm. Recomi
can solve the stated problem of finding optimal perceptrons of the form
(\ref{positive-embeddings})(\ref{and-the-fields}) in finite time, if the
training patterns are in general position, i.e.\ if every subset $\{\xi_\mu\}$
with not more than $N$ elements ($\card(\{\xi_\mu\}) \leq N$) is linearly
independent. It does so in not more than ${\cal O}(N^4)$ floating point
operations. There is no divergence of learning times at the critical storage
capacity $\alpha_c=2$ (for unbiased random patterns), where $\alpha=p/N$.
I am going to show this numerically. In the last section I will
deduce some important constituents of a proof of convergence --- unfortunately
a full proof cannot yet be given. I will analyze there the properties of
locally stable solutions of the optimization problems
(\ref{optim-prob-pos-kappa}) and (\ref{optim-prob-neg-kappa}). It can be shown
that Recomi always stops in a local optimum. If an optimal solution with
$\kappa>0$ exists, Recomi must stop there. Otherwise it is going to stop in one
of the locally stable solutions with $\kappa<0$.

\section*{Description of the algorithm}

Recomi is an iterative algorithm. It calculates coupling vectors $\vect J^{(t)}
= \sum_\mu x_\mu^{(t)} \vect\xi^\mu$ and finds after a finite number of
iterations a solution of the form
(\ref{positive-embeddings})(\ref{and-the-fields}), if it exists. As we will see
later, the algorithm must be initialized with positive embedding strengths
$x_\mu^{(0)}\geq 0$, e.g.\ Hebbian couplings $\vect J^{(0)} = \sum_\mu
\vect\xi^\mu$. For numerical stability $\vect J^{(t)}$ is normalized to 1 after
each iteration. Let $C_\Gamma$ be the correlation matrix of the patterns in
$\Gamma\subseteq\{1,\ldots,p\}$:

\begin{equation}
C_\Gamma = \left( {\vect\xi^\mu}^T \vect\xi^\nu
\right)_{\mu,\nu\in\Gamma}
\end{equation}

\subsection*{Iteration loop}

Let $\vect J^{(t)}$ be given (from now on I drop the index $t$):
\begin{eqnarray}
	\vect J & = & \sum_{\mu=1}^p x_\mu \vect\xi^\mu
	\qquad \left( |\vect J| = 1 \right) \\
	\kappa & = & \min_\mu \left\{ h_\mu \right\} =
        \min_\mu \left\{ \vect J^T \vect\xi^\mu \right\}
\end{eqnarray}

\noindent Let $\Gamma$ be the subset of patterns with minimal local field
$h_\mu$:

\begin{equation}
\Gamma = \left\{ \: \mu \; \big| \; h_\mu = \kappa \: \right\}
\end{equation}

\noindent We now want to alter $\vect J$
\begin{equation}
\vect J \quad \longrightarrow \quad
\vect J' = \sum_{\mu=1}^p \left( x_\mu +
\eps \Delta x_\mu \right) \vect\xi^\mu
\end{equation}
so that for all patterns in $\Gamma$ the local fields grow equally
\begin{equation}
h_\mu' = {\vect J'}^T \vect\xi^\mu = \kappa+\eps
\qquad \forall\mu\in\Gamma.
\end{equation}

\noindent We therefore choose $\Delta\vect x$ to be the pseudo-inverse [9,10]
of the patterns in $\Gamma$:
\begin{equation}
\Delta x_\mu = \cases
{ \sum_{\nu\in\Gamma} \left( C_\Gamma^{-1} \right)_{\mu\nu} &
$\mu\in\Gamma$ \cr \qquad 0 & else \cr }.
\label{def-of-dx}
\end{equation}

If the training patterns $\vect\xi_\mu$ are in general position, $C_\Gamma$
becomes singular if and only if the number of patterns in $\Gamma$,
$\card(\Gamma)$, is greater than $N$. Then Recomi must stop, with
$\vect J^{(t)}$ being the best solution found. Nevertheless Recomi is able to
find optimal solutions as I will show in the last section of this paper.

Now we want to determine the learning rate $\eps$ in a way that {\em all\ }
local fields $h_\mu'$ are greater or equal $\kappa+\eps$:

\begin{equation}
h_\mu' = {\vect J'}^T \vect\xi^\mu \geq \kappa+\eps
\qquad \forall\mu\in\{1,\ldots,p\}.
\label{all_local_fields}
\end{equation}

\noindent $\eps=\eps_\mu$ is the value of the learning rate whith which we get
the equality $h_\mu'=\kappa+\eps$ for pattern $\mu$:
\begin{equation}
\eps_\mu = { h_\mu - \kappa \over 1 -
\sum_{\nu\in\Gamma} C_{\mu\nu} \Delta x_\nu }.
\end{equation}

\noindent To fulfill eqn.~(\ref{all_local_fields}) $\:\eps$ must be smaller or
equal to all relevant, i.e.\ all positive, $\eps_\mu$. We therefore define the
set $\Phi$:
\begin{equation}
\Phi \;\;=\;\; \left\{ \;\; \eps_\mu  \;\; \bigg| \;\;
\mu\notin\Gamma \quad \hbox{and} \quad
0 < \eps_\mu < \infty \;\; \right\}.
\end{equation}

\noindent If $\Phi$ is not empty we can determine $\eps$ as
\begin{equation}
\eps = \min \Phi.
\end{equation}

\noindent If $\Phi$ is empty, we set $\eps=\infty$, i.e. $\vect J' =
\sum_{\mu\in\Gamma} \Delta x_\mu \vect\xi^\mu$, and stop the iteration.

Now $\vect J^{(t+1)} = \vect J' / |\vect J'|$ and we continue at the beginning
of the iteration loop. It is easy to show  that always $\kappa^{(t+1)} =
(\kappa^{(t)}+\eps)/|\vect J'| > \kappa^{(t)}$ (see Appendix). If no solution
with positive $\kappa$ can be found the algorithm typically stops with $\vect
J' = \vect 0$, as will be shown later. (It should be noted that this is the
most sensitive part of the algorithm. Rounding errors must be controlled when
calculating the norm of $\vect J'$.) Then $\vect J^{(t)}$ is taken as the best
solution found by Recomi.

\subsection*{Optimal Recomi}

\begin{figure}
\mypicture{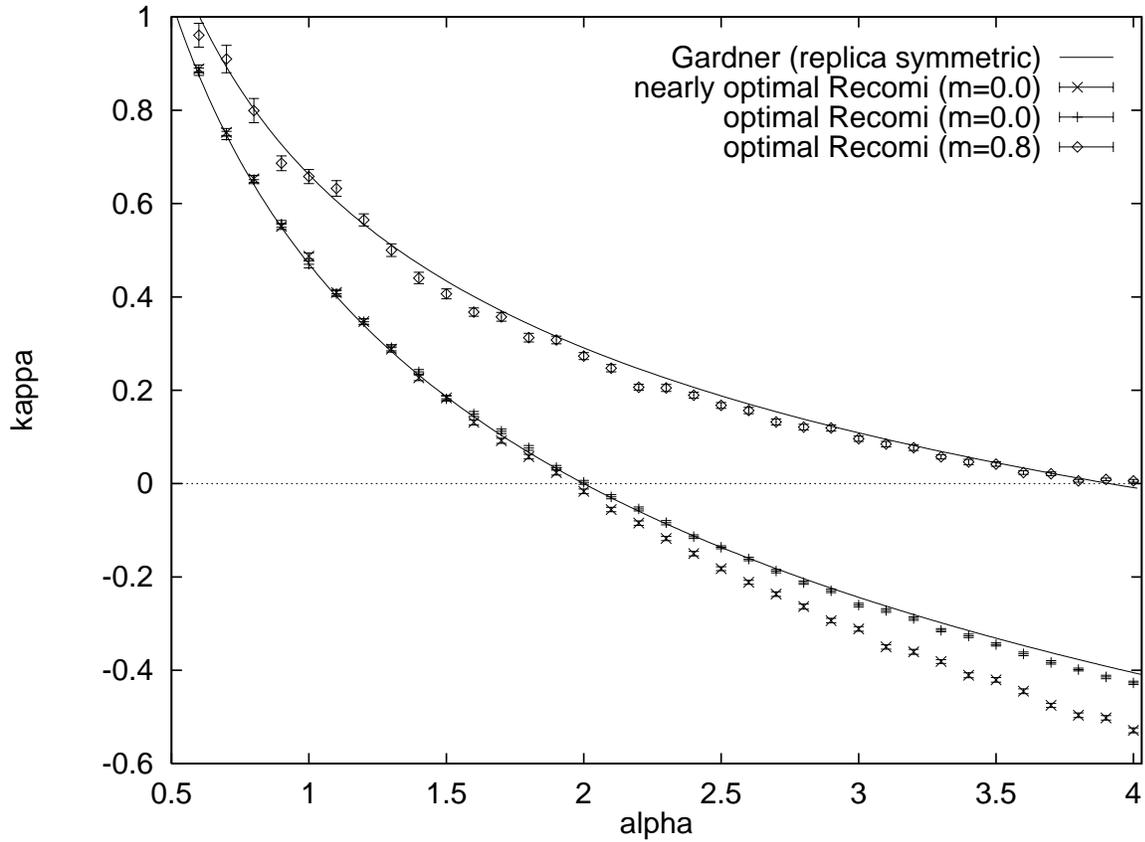}
\caption{\vtop{\hsize=0.88\hsize Comparison of Recomi with Gardner's result,
$N=100$, 100 sets of unbiased ($m=0$) and biased ($m=0.8$) random binary
patterns for each measurement}}
\label{fig1}
\end{figure}

The algorithm I have described so far does not yet find optimal solutions of
the form (\ref{positive-embeddings})(\ref{and-the-fields}). As the changes of
embedding strenghts $\Delta x_\mu$ might be negative in eqn.~(\ref{def-of-dx})
the $x_\mu$ might also become negative in the end. But already this version of
the algorithm does find nearly optimal solutions $\kappa>0$, as can be seen
in fig.~\ref{fig1}, where I compare results for unbiased random patterns
($N=100$) with Gardner's result [3]. Therefore I refer to this version of
Recomi as ``nearly optimal Recomi''.

To find optimal solutions of the form
(\ref{positive-embeddings})(\ref{and-the-fields}) it is necessary to start with
positive embedding strengths $x_\mu\geq 0$, and to make sure that they stay
positive throughout the iteration, i.e.\ $\Delta x_\mu\geq 0$. This is possible
by altering eqn.~(\ref{def-of-dx}). $\Gamma$ must be replaced by a subset
$\Gamma'\subseteq\Gamma$ with the following properties:

\begin{equation}
\Gamma'\subseteq\Gamma
\end{equation}
\begin{equation}
\Delta x_\mu = \sum_{\nu\in\Gamma'} \left( C_{\Gamma'}^{-1}
\right)_{\mu\nu} \geq 0 \qquad \forall\mu\in\Gamma'
\end{equation}
\begin{equation}
\left( \sum_{\nu\in\Gamma'} \Delta x_\nu \vect\xi^\nu \right)^T
\vect\xi^\mu \geq 1 \qquad \forall\mu\in\Gamma
\end{equation}

It is always possible to find such a subset $\Gamma'$ (as long as $C_\Gamma$
itself is regular), because $\sum_{\nu\in\Gamma'} \Delta x_\nu \vect\xi^\nu$
then is the (unique) optimal perceptron for the correct mapping of the patterns
$\mu\in\Gamma$.

$\Gamma'$ can easily be determined. The following algorithm
proved to work in all cases tested (about ${\cal O}(10^5)$ algorithm runs). I
cannot yet prove its convergence analytically. This has to be done in later
work. To find $\Gamma'$ one can proceed as follows:

1.)\ \ \vtop{\hsize=0.85\hsize\noindent start with $\Gamma'=\Gamma$}

2.)\ \ \vtop{\hsize=0.85\hsize\noindent calculate $\Delta x_\mu =
\sum_{\nu\in\Gamma'} \left( C_{\Gamma'}^{-1} \right)_{\mu\nu} (\mu\in\Gamma')$;
$\Delta x_\varrho=\min_\mu \{\Delta x_\mu\}$; if $\Delta x_\varrho < 0$ remove
$\varrho$ from $\Gamma'$ and go to 2.) else go to 3.)}

\medskip

3.)\ \ \vtop{\hsize=0.85\hsize\noindent calculate $\Delta h_\mu = (
\sum_{\nu\in\Gamma'} \Delta x_\nu \vect\xi^\nu )^T \vect\xi^\mu
\:(\mu\in\Gamma\setminus\Gamma')$; $\Delta h_\sigma=\min_\mu \{\Delta h_\mu\}$;
if $\Delta h_\sigma < 1$ add $\sigma$ to $\Gamma'$ and go to 2.) else STOP}

\medskip

\begin{figure}
\mypicture{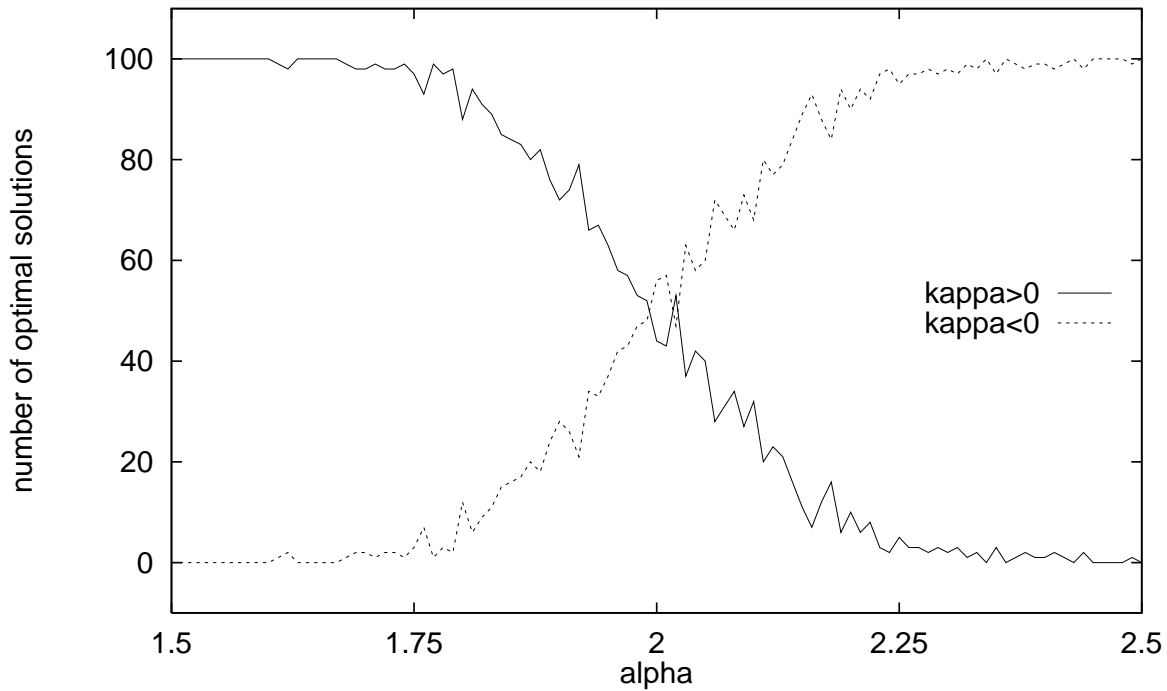}
\caption{\vtop{\hsize=0.88\hsize Optimal Recomi, $N=100$, 100 sets of
unbiased random binary patterns for each value of $\alpha$}}
\label{fig2}
\end{figure}

By replacing $\Gamma$ by $\Gamma'$ in eqn.~(\ref{def-of-dx}) Recomi is able to
find optimal solutions. I refer to this improved version of the algorithm as
``optimal Recomi''. In fig.~\ref{fig2} I check for unbiased random binary
patterns ($N=100$), how often the algorithm stops in optimal solutions with
$\kappa>0$, and in locally optimal solutions with $\kappa<0$. For every value
of $\alpha=p/N$ 100 different pattern sets are tested. In very rare cases
(not in this figure) the algorithm only gets close to but does not reach
optimal solutions: trying to invert nearly singular correlation matrices can
cause failure of the inversion subroutines.

In fig.~\ref{fig1} I compare results for unbiased and biased
random binary patterns with Gardner's result [3]. The patterns $\eta^\mu_i$
are chosen with a probability distribution $p(\eta^\mu_i) = {(1-m)\over 2}
\,\delta(\eta^\mu_i+1) + {(1+m)\over 2} \,\delta(\eta^\mu_i-1)$,
using $m=0$ (unbiased) and $m=0.8$ (biased), and the
$\xi^\mu_i$ calculated according to eqn.~(\ref{xi}). Within the error bounds
there is no difference to be seen between optimal and nearly optimal solutions
below $\alpha_c$ ($\kappa>0$). In the range of replica symmetry breaking
$\alpha>\alpha_c$ ($\kappa<0$) optimal Recomi clearly performs better than the
simpler version of the algorithm. Here it cannot be expected that the algorithm
finds a global stability optimum, as it gets trapped in one of the many local
optima, which will be shown in the last section of this paper. Note that for
the biased patterns ($m=0.8$) at $N=100$ one still has to take finite size
effects into account: the measured points are
all optimal solutions, but yet still lie a little bit below
the Gardner curve. Also note that the theoretical lines are all calculated in
replica symmetric approximation, i.e.\ they must be corrected for negative
$\kappa$, where replica symmetry is no longer valid.

\begin{figure}
\mypicture{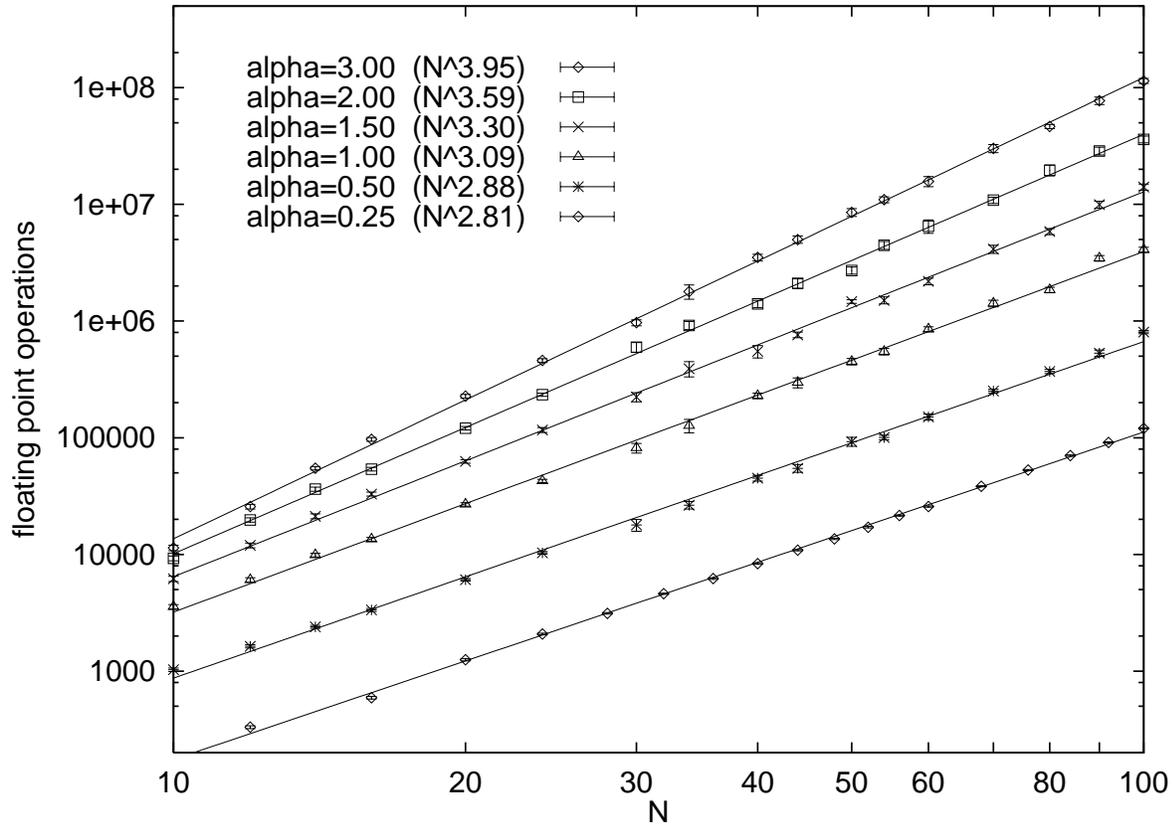}
\caption{\vtop{\hsize=0.88\hsize Convergence time against system size for
optimal Recomi (unbiased random binary patterns)}}
\label{fig3}
\end{figure}

In fig.~\ref{fig3} I train perceptrons of different sizes $N$ with unbiased
random binary patterns. Convergence time is plotted against system size $N$ for
different values of the storage capacity $\alpha$. The most expensive part of
the algorithm, in the large $N$ limit, is matrix inversion, which is of ${\cal
O}(N^3)$ for each single inversion. Nearly optimal Recomi therefore is, in the
worst case, of ${\cal O}(\sum_{i=1}^N i^3)={\cal O}(N^4)$, as $\card(\Gamma)$
grows at least by one in each iteration step. For optimal Recomi one cannot
give such a simple derivation of convergence times, as $\card(\Gamma)$ can also
shrink in the learning process. But here convergence time is also bounded from
above by ${\cal O}(N^4)$: In fig.~\ref{fig3} I count the number of floating
point operations ($+-*/$) optimal Recomi needs
to find solutions. As below $N=100$
convergence time is still dominated by other operations apart from matrix
inversion, I only plot the matrix inversion part here. All other operations
are of ${\cal O}(N^3)$ or below. Just as predicted for nearly optimal Recomi
the optimal version of the algorithm converges in ${\cal O}(N^4)$ or less
floating point operations.

\begin{figure}
\mypicture{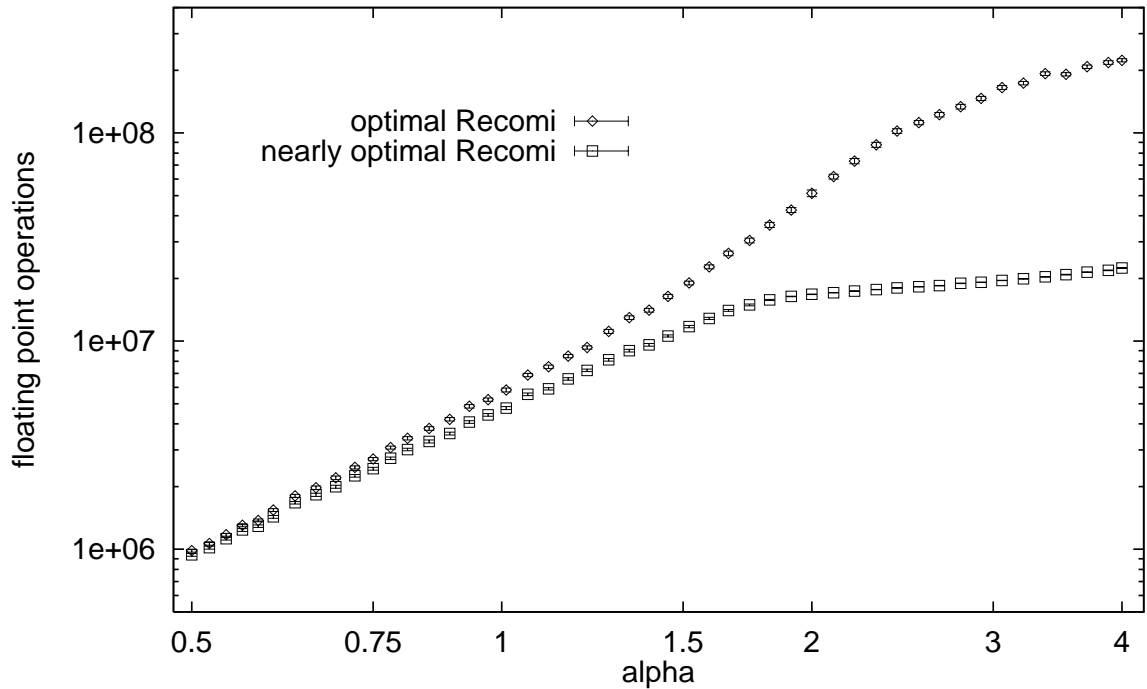}
\caption{\vtop{\hsize=0.88\hsize Convergence time against storage capacity
$\alpha$ (unbiased random binary patterns, $N=100$). There is no
divergence at $\alpha_c=2$.}}
\label{fig4}
\end{figure}

In fig.~\ref{fig4} I plot convergence time (i.e.\ number of floating
point operations) against the storage capacity $\alpha$. Again the perceptron
($N=100$) was trained with unbiased random binary patterns. There is no
divergence at $\alpha=\alpha_c=2$. For small $\alpha$ the two versions of
the algorithm differ only little, as nearly optimal Recomi also often
finds optimal solutions (see also fig.~\ref{fig1}). For larger values
of $\alpha$ the convergence times evolve different.

\section*{Analysis of local stability optima: towards a proof of convergence}

I cannot yet give a full proof of convergence of Recomi, but some major
components can already be deduced. For this reason I want to consider the role
of local stability optima. It is useful here to use the problem formulations
eqn.~(\ref{optim-prob-pos-kappa}) (for $\kappa>0$) and
eqn.~(\ref{optim-prob-neg-kappa}) (for $\kappa<0$). If I write a $\pm$-sign in
the following text, the $+$ always refers to the case $\kappa>0$ and the $-$ to
$\kappa<0$.

The Problem (\ref{optim-prob-pos-kappa})(\ref{optim-prob-neg-kappa}) can now be
formulated as
\begin{eqnarray}
\hbox{minimize} & f(\vect x) = \pm \; \vect J^T \vect J =
\pm \; \vect x^T C \vect x & \nonumber \\
\hbox{under the constraints} & h_\mu = \vect J^T \vect\xi^\mu =
(C \vect x)_\mu \geq \pm 1 & \quad \mu = 1,\ldots,p.
\label{optim-prob}
\end{eqnarray}

$\Gamma$ is the set of patterns with minimal local field:
\begin{equation}
\Gamma = \left\{ \: \mu \; \big| \; h_\mu=(C \vect x)_\mu=\pm 1 \: \right\}.
\end{equation}

Let $\Omega$ be the set of all possible search directions $\Delta\vect x$,
which do not violate the inequality constraints eqn.~(\ref{optim-prob}):

\begin{equation}
\Omega = \left\{ \: \Delta\vect x \in \real^p \; \bigg| \; (C\Delta\vect x)_\mu
\geq 0 \;\;\;\; \forall\mu\in\Gamma \: \right\}
\end{equation}

A solution $\vect J$ is locally optimal if and only if
\begin{equation}
\Big[ \nabla_x f(\vect x) \Big]^T \Delta\vect x =
\pm\; 2 \: \vect x^T C \Delta\vect x
\geq 0 \qquad \forall \Delta\vect x \in \Omega
\label{local-opt}
\end{equation}

I now prove the important theorem, that if there is a solution with positive
stability $\kappa>0$ there cannot be locally stable solutions $\vect J$ with
negative stability $\kappa<0$ and $x_\mu\geq 0$ $\forall\mu$, $\sum_\mu
x_\mu>0$:

If there is a solution with $\kappa>0$ there must be a solution of the form
(e.g.\ the optimal perceptron)
\begin{equation}
\vect J^\ast = \sum_\mu x_\mu^\ast \vect\xi^\mu, \quad\hbox{with\ \ }
{\vect J^\ast}^T \vect\xi^\mu = (C\vect x^\ast)_\mu \geq\kappa^\ast >0
\quad\forall\mu
\end{equation}

\noindent Let us assume $\vect J = \sum_\mu x_\mu \vect\xi^\mu$ is locally
optimal with $\kappa = \min_\mu \{\vect J^T \vect\xi^\mu\} < 0$ and $x_\mu\geq
0$ $\forall\mu$, $\sum_\mu x_\mu>0$. That means (eqn.~(\ref{local-opt})):
\begin{equation}
\vect x^T C \Delta\vect x \leq 0 \qquad \forall \Delta\vect x \in \Omega
\label{contra}
\end{equation}

\noindent As $(C\vect x^\ast)_\mu\geq\kappa^\ast>0$ $\forall\mu$, we have:
\begin{equation}
\Delta\vect x \definition \vect x^\ast \in \Omega
\end{equation}
\begin{equation}
\vect x^T C \Delta\vect x = \vect x^T C \vect x^\ast =
\sum_\mu x_\mu (C\vect x^\ast)_\mu \geq \kappa^\ast \sum_\mu x_\mu > 0
\end{equation}

\noindent in contradiction to eqn.~(\ref{contra})! Therefore such a vector
$\vect J$ cannot exist. We will see below that optimal Recomi always stops in
(local) optima which by definition of the algorithm are of the form $x_\mu\geq
0$ $\forall\mu$ and $\sum_\mu x_\mu>0$. So if there is any solution with
$\kappa>0$ Recomi can only stop in the global optimum of the problem, because
then there are no other optima of that form.

To show this, I have to make several assumptions, which I cannot prove yet: a)
The algorithm described in section ``Optimal Recomi'' for deriving $\Gamma'$
really always works. b) The size of $\Gamma$, $\card(\Gamma)$, grows not more
than by one in each iteration step, especially not from $\card(\Gamma)<N$ to
$\card(\Gamma)>N$. c) Recomi really terminates in finite time. About this last
point one can only say that $\kappa^{(t)}$ is a strictly monotonical function
of $t$ (see Appendix), i.e.\ there is always an attractor of the training
dynamics.

If these three assumptions are correct, Recomi stops in a (local) optimum,
which is the global one, if solutions $\kappa>0$ exist. To show this I have to
consider the three possible ways the algorithm does stop: 1) $\Phi$ is empty,
i.e.\ $\eps$ becomes infinit. 2) $\vect J'$ is zero. 3) $C_\Gamma$ is singular.

1) $\Phi$ is empty: This is the most simple case. Then, by definition, $\vect
J' = \sum_{\mu\in\Gamma'} \Delta x_\mu \vect\xi^\mu$, which is an optimal
solution of the form (\ref{positive-embeddings})(\ref{and-the-fields}). This is
the usual way Recomi stops if solutions $\kappa>0$ exist.

2) $\vect J'$ is zero: Then $\vect J^{(t)}=-\eps \sum_{\mu\in\Gamma'} \Delta
x_\mu \vect\xi^\mu$. Applying the Kuhn-Tucker theorem this is a locally stable
solution for $\kappa<0$ (just like
(\ref{positive-embeddings})(\ref{and-the-fields}) for $\kappa>0$). As $\vect
J^{(t)}$ is coded in the form $x_\mu\geq 0$ $\forall\mu$ and $\sum_\mu x_\mu>0$
there cannot be solutions with $\kappa>0$ as was shown above. This is the usual
way Recomi stops if no solutions $\kappa>0$ exist.

3) $C_\Gamma$ is singular: Then $\card(\Gamma)>N$ (because the training
patterns are in general position). According to our assumption, $\card(\Gamma)$
must have been $N$ in the iteration step before. $\Gamma'$ must have been equal
to $\Gamma$ because otherwise $\card(\Gamma)$ would not have grown. As $\{
\xi^\mu | \mu\in\Gamma \}$ does span $\real^N$, $\vect J^{(t-1)}$ is completely
determined by the local fields ${\vect J^{(t-1)}}^T \vect\xi^\mu$
$\mu\in\Gamma$, i.e. $\vect J^{(t-1)} \sim \sum_{\mu\in\Gamma} \Delta x_\mu
\vect\xi^\mu$, which is a local optimum. Therefore case 3) does in principal
never occur, the algorithm stops before in 1) or 2).

In practice case 3) does occur, as sometimes nearly singular correlation
matrices cannot be inverted by the inversion subroutines because of numerical
restrictions.

\section*{Conclusion}
In this article I presented a perceptron learning algorithm, which is able to
find the optimal perceptron in finite time, i.e.~in ${\cal O}(N^4)$
floating point operations. The algorithm even works beyond the
critical storage capacity $\alpha_c$, where it finds solutions of negative
stability that are locally optimal. Calculating the stability curve
$\kappa(\alpha)$ for random training patterns exactly reproduces Gardner's
predictions [3]. A full prove of convergence could not yet be given, but major
constituents were already shown. As the algorithm works very reliably, it can
be expected that a full proof of convergence can be found. Furthermore it is
planned to generalize the algorithm to two layer perceptrons with fixed output.
First results are very promising, yet it cannot be expected that the algorithm
finds globally optimal solutions, because replica symmetry breaking effects
are very strong in this case.

\section*{Appendix}

In this appendix I will show that $\kappa^{(t)}$ is a strictly monotonical
function of $t$:
\begin{equation}
\kappa^{(t+1)} = { \kappa^{(t)} + \eps \over |\vect J'| }
\end{equation}
\begin{equation}
\varrho \definition \left( \sum_\mu \Delta x_\mu \vect\xi^\mu \right)^2 =
\sum_{\mu\nu} \Delta x_\mu (C_\Gamma)_{\mu\nu} \Delta x_\nu =
\sum_{\mu\nu\lambda} \Delta x_\mu (C_\Gamma)_{\mu\nu}
(C^{-1}_\Gamma)_{\nu\lambda} =
\sum_\mu \Delta x_\mu \geq 0
\end{equation}
\begin{equation}
{\vect J'}^T \vect J' = \left( \vect J^{(t)} +
    \eps \sum_{\mu\in\Gamma} \Delta x_\mu \vect\xi^\mu \right)^2 =
    1 + 2 \eps \kappa^{(t)} \varrho + \eps^2 \varrho
    \geq 0 \qquad\forall\eps\in\real
\end{equation}
\begin{equation}
\hbox{e.g.} \quad \eps = -\kappa^{(t)} \quad\Longrightarrow\quad
1 - {\kappa^{(t)}}^2 \varrho \geq 0
\end{equation}
\begin{equation}
{d\over d\eps} \kappa^{(t+1)} =
\left(1 + 2 \eps \kappa^{(t)} \varrho + \eps^2 \varrho\right)^{-3/2}
\left(1 - {\kappa^{(t)}}^2 \varrho\right) \geq 0
\end{equation}

\noindent ${d\over d\eps} \kappa^{(t+1)} = 0$ if and only if Recomi stops in a
(local) optimum:
\begin{equation}
{d\over d\eps} \kappa^{(t+1)} = 0 \iff 1 - {\kappa^{(t)}}^2 \varrho = 0
\iff \vect J^{(t)} = \kappa^{(t)} \sum_{\mu\in\Gamma} \Delta x_\mu \vect\xi^\mu
\end{equation}

\noindent That means $\kappa^{(t+1)}>\kappa^{(t)}$ as long as Recomi has not
terminated. qed.

\section*{Acknowledgements}
I would like to thank Prof.\ H.\ Horner, Dr.\ R.\ K\"uhn and Prof.\ H.\ G.\
Bock for valuable discussions. This work was sponsored by the IWR
(Interdisciplinary Center for Scientific Computing), Universit\"at Heidelberg.

\section*{References}

\reference{1}{J.J.\ Hopfield}{Neural Networks and Physical Systems with
Emergent Computational Abilities}{Proc.\ Natl.\ Acad.\ Sci.\ USA {\bf 79},
2554--2559 (1982)}
\reference{2}{D.J.\ Amit, H.\ Gutfreund and H.\ Sompolinsky}{Storing Infinite
Numbers of Patterns in a Spin-Glass Model of Neural Networks}{Phys.\ Rev.\
Lett.\ {\bf 55}, 1530--1533 (1985)}
\reference{3}{E.\ Gardner}{The Space of Interactions in Neural Network Models}
{J.\ Phys.\ A {\bf 21}, 257--270 (1988)}
\reference{4}{S.\ Diederich and M.\ Opper}{Learning of Correlated Patterns in
Spin-Glass Networks by Local Learning Rules}{Phys.\ Rev.\ Lett.\ {\bf 58} (9),
949--952 (1987)}
\reference{5}{W.\ Krauth and M.\ M\'ezard}{Learning algorithms with optimal
stability in neural networks}{J.\ Phys.\ A {\bf 20}, L745--L752 (1987)}
\reference{6}{J.\ K.\ Anlauf and M.\ Biehl}{The AdaTron: an Adaptive Perceptron
Algorithm}{Europhys.\ Lett.\ {\bf 10} (7), 687--692 (1989)}
\reference{7}{M.\ Opper}{Learning times of neural networks: Exact solution for
a PERCEPTRON algorithm}{Phys.\ Rev.\ A {\bf 38} (7), 3824--3826 (1988)}
\reference{8}{P.\ Ruj\'an}{A fast method for calculating the perceptron with
maximal stability}{J.\ Physique I {\bf 3}, 277--290 (1993)}
\reference{9}{L.\ Personnaz, I.\ Guyon and G.\ Dreyfus}{Information storage and
retrieval in spin-glass like neural networks}{J.\ Physique Lett.\ {\bf 46},
L359--L365 (1985)}
\reference{10}{I.\ Kanter and H.\ Sompolinsky}{Associative recall of memory
without errors}{Phys.\ Rev.\ A {\bf 35} (1), 380--392 (1987)}
\reference{11}{R.\ Fletcher}{Practical Methods of Optimization}{Wiley, New York
(1987)}

\newpage

\end{document}